\documentclass[12pt,preprint]{aastex}
\usepackage{emulateapj5}
\slugcomment{To appear in ApJ}

\begin{document}
\title{Shape Reconstruction and Weak Lensing Measurement
with Interferometers:  A Shapelet Approach} 
\author{Tzu-Ching Chang$^{1}$ and Alexandre Refregier$^{2}$} 
\vspace{0.1cm}
\affil{1 Department of Astronomy and Columbia Astrophysics Laboratory, 
Columbia University, 550 W. 120th Street, New York, NY 10027, USA; 
tchang@astro.columbia.edu}
\affil{2 Institute of Astronomy, Madingley Road, Cambridge CB3 0HA,
UK; ar@ast.cam.ac.uk}

\begin{abstract}
We present a new approach for image reconstruction and weak lensing
measurements with interferometers.  Based on the shapelet formalism
presented in Refregier (2001), object images are decomposed into
orthonormal Hermite basis functions. The shapelet coefficients of a
collection of sources are simultaneously fit on the $uv$ plane, the
Fourier transform of the sky brightness distribution observed by
interferometers. The resulting $\chi^2$-fit is linear in its
parameters and can thus be performed efficiently by simple matrix
multiplications. We show how the complex effects of bandwidth
smearing, time averaging and non-coplanarity of the array can be
easily and fully corrected for in our method.  Optimal image
reconstruction, co-addition, astrometry, and photometry can all be
achieved using weighted sums of the derived coefficients.  As an
example we consider the observing conditions of the FIRST radio survey
(Becker, White \& Helfand 1995; White et al. 1997). We find that our
method accurately recovers the shapes of simulated images even for the
sparse $uv$ sampling of this snapshot survey.  Using one of the FIRST
pointings, we find our method compares well with CLEAN, the commonly
used method for interferometric imaging.  Our method has the advantage
of being linear in the fit parameters, of fitting all sources
simultaneously, and of providing the full covariance matrix of the
coefficients, which allows us to quantify the errors and cross-talk in
image shapes.  It is therefore well-suited for quantitative shape
measurements which require high-precision. In particular, we show how
our method can be combined with the results of Refregier \& Bacon
(2001) to provide an accurate measurement of weak lensing from
interferometric data.
\end{abstract}

\section{Introduction}
Interferometers are widely used for astronomical observations as they
provide high angular resolutions and large collecting areas. Existing
interferometers in the radio (e.g. the Very Large Array (VLA), the
Berkeley-Illinois-Maryland Association (BIMA), etc.) and in the
optical band (e.g., the Center for High Angular Resolution Astronomy
(CHARA), the Palomar Testbed Interferometer, the Keck Interferometer,
the Very Large Telescope (VLT) Interferometer, etc.) will soon be
complemented by new facilities such as the Expanded Very Large Array
(EVLA), the Square Kilometer Array (SKA), the Atacama Large Millimeter
Array (ALMA), and the Low Frequency Array (LOFAR). Interferometers
are now also being developed to produce maps of the Cosmic Microwave
Background on small scales (e.g., the Cosmic Background Imager (CBI),
the Array for Microwave Background Anisotropy (AMiBA), the Very Small
Array (VSA), etc).

Interferometric arrays, however, do not provide a direct image of the
observed sky, but instead measure its Fourier transform at a finite
number of discrete samplings, or `$uv$' points, corresponding to each
antenna pair in the array. The image in real space must therefore be
reconstructed from the $uv$ plane by inverse Fourier transform while
deconvolving the effective beam arising from the finite sampling (see
Thompson et al. 1986, Perley et al. 1989, and Taylor et al. 1999 for
reviews). For this purpose several elaborate methods have been
developed.  For instance, the commonly used CLEANing algorithm
implemented in the NRAO AIPS software package (Hogbom 1974; Schwarz
1978; Clark 1980; Cornwell 1983), relies on successive subtraction of
real-space delta functions from the $uv$ plane. Another method is
based on Maximum Entropy (e.g., Cornwell \& Evans 1985) and consists
of finding the simplest image consistent with the $uv$ data. These
methods are well-tested and appropriate for various applications;
however, the methods are non-linear and do not necessarily converge in
a well-defined manner. Consequently, they are not well-suited for
quantitative image shape measurements requiring high precision. In
particular, weak gravitational lensing (see Mellier 1999; Bartelmann
\& Schneider 2000 for reviews) requires the statistical measurements
of weak distortions in the shapes of background objects and thus
cannot afford the instabilities and potential biases inherent in these
methods. While interferometric surveys offer great promises for weak
lensing (Kamionkowski et al. 1998; Refregier et al 1998;
Schneider 1999), a different approach for shape measurements is therefore
required to achieve the necessary accuracy and control of systematics.

In this paper, we present a new method for reconstructing images from
interferometric observations. It is based on the formalism introduced
by Refregier (2001, Paper I) and Refregier \& Bacon (2001, Paper II),
in which object shapes are decomposed into orthonormal shape
components, or `shapelets'.  The Hermite basis functions used in this
approach have a number of remarkable properties which greatly
facilitate the modeling of object shapes. In particular, they are
invariant under Fourier transformation (up to a rescaling) and are
thus a natural choice for interferometric imaging. We show how
shapelet components can be directly fitted on the $uv$ plane to
reconstruct an interferometric image. The fit is linear in the
shapelet coefficients and can thus be performed by simple matrix
multiplications. Since the shapelet components of all sources are
fitted simultaneously, cross-talk between different sources (e.g.,
when the sidelobe from one source falls at the position of a second
source) are avoided, or at least quantified.  The method also provides
the full covariance matrix of the shapelet coefficients, and is
robust. We also show how the complex effects of bandwidth smearing,
time averaging and non-coplanarity of the array can be easily and
fully corrected for in our method. Our method is thus well-suited for
applications requiring unbiased, high-precision measurements of object
shapes. In particular, we show how the method can be combined with the
results of Paper II to provide a clean measurement of weak
gravitational lensing with interferometers. We test our methods using
both observations from the FIRST radio survey (Becker et al. 1995;
White et al. 1997) and numerical simulations corresponding to the
observing conditions of that survey. We also show how our method can
be implemented on parallel computers and discuss its performance in
comparison with the CLEANing method.

Our paper is organized as follows. In \S\ref{shapelets}, we first
summarize the relevant features of the shapelet method. In
\S\ref{method}, we describe how shapelets can be applied to image
reconstruction with interferometers. In \S\ref{results}, we discuss
tests of the method using both simulated and real FIRST observations. In
\S\ref{lensing} we show how our method can be used for weak lensing
applications. Our conclusions are summarized in \S\ref{conclusion}.

\section{Shapelet Method}
\label{shapelets}
We begin by summarizing the relevant components of the shapelet method
described in Paper I. In this approach, the surface brightness $f({\mathbf
x})$ of an object is decomposed as
\begin{equation}
f({\mathbf x}) = \sum_{{\mathbf n}} f_{{\mathbf n}} B_{{\mathbf n}}
({\mathbf x};\beta),
\end{equation}
where
\begin{equation}
B_{{\mathbf n}}({\mathbf x};\beta) \equiv 
\frac{H_{n_{1}}(\beta^{-1} x_{1}) ~H_{n_{2}}(\beta^{-1} x_{2})
   ~e^{-\frac{|x|^{2}}{2 \beta^{2}}}}
  {\left[ 2^{(n_{1}+n_{2})} \pi ~\beta^{2}
  ~n_{1}! ~n_{2}! \right]^{\frac{1}{2}}}
\end{equation}
are the two-dimensional orthonormal Hermite basis functions of
characteristic scale $\beta$, $H_{m}(\xi)$ is the Hermite polynomial
of order m, ${\mathbf x}=(x_{1},x_{2})$ and ${\mathbf
n}=(n_{1},n_{2})$. The basis is complete and yields fast convergence
in the expansion if $\beta$ and $x=0$ are, respectively, close to the
size and location of the object.  The basis functions can be thought
of as perturbations around a two-dimensional Gaussian, and are thus
natural bases for describing the shapes of most astronomical
objects. They are also the eigenfunctions of the Quantum Harmonic
Oscillator (QHO), allowing us to use the powerful formalism developed
for that problem.  A similar decomposition scheme using Laguerre basis
functions has been independently proposed by Bernstein \& Jarvis
(2001).

The Hermite basis functions have remarkable mathematical properties. In
particular, let us consider the Fourier transform of an object
intensity, $\tilde{f}({\mathbf k}) = (2\pi)^{-\frac{1}{2}}
\int_{-\infty}^{\infty} d^{2}{x} f({\mathbf x}) e^{i{\mathbf k} \cdot
{\mathbf x}}$. It can be decomposed as $\tilde{f}({\mathbf k}) =
\sum_{{\mathbf n}} f_{{\mathbf n}} ~\widetilde{B}_{{\mathbf
n}}({\mathbf k};\beta)$, where $\widetilde{B}_{{\mathbf n}}({\mathbf
k};\beta)$ are the Fourier-transforms of the basis functions, which
obey the dual property
\begin{equation}
\label{eq:duality}
\widetilde{B}_{{\mathbf n}}({\mathbf k};\beta) = i^{(n_{1}+n_{2})}
B_{{\mathbf n}}({\mathbf k};\beta^{-1}).
\end{equation}
From the orthonomality of the basis functions, the coefficients are
given by
\begin{equation}
f_{{\mathbf n}} = \int_{-\infty}^{\infty} ~d^{2}{k}~ \tilde{f}
 ({\mathbf k}) ~\widetilde{B}_{{\mathbf n}}({\mathbf k};\beta).
\end{equation}
This invariance (up to a rescaling) under Fourier transformation 
(Eq.~[\ref{eq:duality}]) makes this basis set a natural choice for
interferometric imaging.

\section{Shapelet Reconstruction with Interferometers}
\label{method}
In this section, we describe how shapelets can be applied to
interferometric imaging. We first briefly discuss how images are
mapped onto the $uv$ plane by interferometers. We also show how the $uv$
plane can be binned into cells to reduce computation time and memory
requirements. We then describe how the shapelet coefficients can be
directly fit onto the binned $uv$ plane using a linear $\chi^2$
procedure. Finally, we describe how the resulting shapelet
coefficients can be optimally combined to reconstruct the image, to
co-add several pointings, and to measure shape parameters.

\subsection{Interferometric Observations}
\label{interferometers}
An interferometer consists of an array of antennae whose output
signals are correlated to measure a complex `visibility' for each
antenna pair (see Thompson et al. 1986, Perley et al. 1989, and Taylor
et al. 1999 for reviews). Each visibility is then assigned a point on
the `$uv$ plane' corresponding to the two-dimensional spacings between
the antennae. In practice, the visibilities are close to, but not
exactly equal to a two-dimensional Fourier transform of the sky
brightness.  Within the conventions of Perley, Schwab \& Bridle (1989)
for the VLA, the visibility measured for the antenna pair $(i,j)$ at
time $t$ and at frequency $\nu$ is indeed given by
\begin{equation}
\label{eq:v_ij}
V_{ij}(\nu,t) = \int d^{2}l~ \frac{A({\mathbf l},\nu)
f({\mathbf l},\nu,t)}{\sqrt{1-|l|^2}} e^{-2\pi i [ ul + vm +w (\sqrt{1-|l|^2}-1)]},
\end{equation}
where $f({\mathbf l},\nu,t)$ is the surface brightness of the sky at
position ${\mathbf l}=(l,m)$ with respect to the phase center, and
$A({\mathbf l},\nu)$ is the (frequency-dependent) primary beam. For
the VLA, the primary beam power pattern can be well-approximated as
the Bessel function $2J_{1}(z)/z$, where $z \simeq 3.234~r
\theta_{p}^{-1}, ~\theta_{p}=30'.83 \times \left(\frac{1.4 {\rm
GHz}}{\nu} \right)$, $\nu$ is the observation frequency and $r$ is the
position offset from the phase center (Condon et al. 1998).  The
$u,v,w$ coordinates are given by
\begin{eqnarray}
\left( \begin{array}{c} u \\ v \\ w \end{array} \right) & = &
\left( \begin{array}{ccc} \sin H_{0} & \cos H_{0} &
0 \\ - \sin \delta_{0} \cos H_{0} & \sin \delta_{0} \sin H_{0} & \cos
\delta_{0} \\ \cos \delta_{0} \cos H_{0} & - \cos \delta_{0} \sin
H_{0} & \sin \delta_{0} \\ \end{array} \right) \nonumber \\
 & & \times \left( \begin{array}{c}
L_{x} \\ L_{y} \\ L_{z} \end{array} \right) \frac{1}{\lambda},
\end{eqnarray}
where $\lambda=c \nu^{-1}$ is the wavelength of observation, $H_{0}$
and $\delta_{0}$ are the hour angle and declination of the phase
center, and $(L_{x},L_{y},L_{z})$ are the coordinate differences for
the two antennas. The latter are measured in a fixed-Earth coordinate
system, for which the sky rotates about the $\hat{L}_{z}$ axis.  Note
that the $(u,v,w)$ positions of the visibilities define the
synthesized beam pattern. Since the $(u,v,w)$ coordinates are entirely
determined by the antenna positions, source coordinates, and time and
frequency of the observations, the synthesized beam is precisely known
for interforemeters.

Only in the absence of a primary beam ($A=1$), for observations at
zenith ($w=0$), and for small displacements from the phase center
($l,m \ll 1$), does the visibility reduce to an exact Fourier
Transform of the intensity.  Furthermore, the visibilities are
measured in practice by averaging over small time and frequency
intervals. The resulting averaged visibility is given by
\begin{equation}
\label{eq:v_ij_bar}
\overline{V}_{ij} = \int dt \int d\nu ~T(t) G(\nu) V_{ij}(t,\nu)
\end{equation}
where $T(t)$ and $G(t)$ are the time and frequency window functions,
respectively, and are normalized as $\int dt T(t) = \int d\nu G(\nu)
\equiv 1$. Because the time and frequency intervals are typically very
small, this double integral can be evaluated by Taylor expanding
$V_{ij}(t,\nu)$ about the central values $t_{0}$ and $\nu_{0}$ of the
window functions. For square-hat window functions of width $\Delta t$
(exact) and $\Delta \nu$ (approximate), respectively, we obtain
\begin{eqnarray}
\overline{V}_{ij} & \simeq & V_{ij}(t_{0},\nu_{0}) + \frac{1}{24} \left[
\frac{\partial^{2} V_{ij}(t_{0},\nu_{0})}{\partial t^{2}} (\Delta
t)^{2} \nonumber \right. \\
 & & \left. + \frac{\partial^{2} V_{ij}(t_{0},\nu_{0})}{\partial \nu^{2}}
(\Delta \nu)^{2} \right] + \cdots
\end{eqnarray}
When the telescope points to a fixed location on the sky, the hour
angle of the phase center changes as $H_{0}(t)\propto \omega_{E}t$,
where $\omega_{E}$ is the angular frequency of the Earth. On the other
hand, the declination $\delta_{0}$ of the phase center remains
constant.

The above expression for $\overline{V}_{ij}$ can thus be computed
analytically, leaving the two-dimensional ${\mathbf l}$-integral to
evaluate numerically. Note that this provides a direct and complete
treatment of primary beam attenuation, time-averaging, bandwidth
smearing and non-coplanarity of the array. These effects are difficult
to correct for in the context of the standard CLEANing method.

\subsection{Binning in the $uv$ plane}
In practice, the number of visibilities per observation is large ($>
10^5$). Directly fitting the shape parameters to all $uv$ points would
thus require prohibitively large computing time and memory.  Instead,
we use a binning scheme to reduce the effective number of $uv$ points
without loosing information. In the $uv$ plane, we set a grid of size
$\Delta u = \frac{1}{2} ~{\Delta l}^{-1}$ and average the visibilities
inside each cell, where $\Delta l$ is one-half of the intended field
of view, and the factor $\frac{1}{2}$ accounts for the Nyquist
frequency.  The choice of $\Delta u$ is designed both to minimize the
number of cells and to avoid smearing at large angular scales, which
would otherwise act like an effective primary beam attenuation.  We
thus calculate the average visibility in the $uv$ cell $c$ (of size
$\Delta u$) as
\begin{equation}
\label{eq:v_c}
\overline{V}_{c} = \frac{1}{N_{c}}\sum_{ij \in c} \overline{V}_{ij}
\end{equation}
where $N_{c}$ is the number of visibilities in the cell. This is the
data we will use to reconstruct the image.

\subsection{Fitting for the shapelet coefficients}
We now wish to model the intensity $f_{s}(l,m)$ of each source $s$
as a sum of shapelet basis functions
\begin{equation}
f_{s}({\mathbf l}) = \sum_{\mathbf n} f_{{\mathbf n}s} B_{\mathbf
n}({\mathbf l}-{\mathbf l}_{s};\beta_{s}),
\end{equation}
centered on the source centroid ${\mathbf l}_s=(l_{s},m_{s})$, and
scale $\beta_{s}$. Our goal is to estimate the shapelet coefficients
$f_{{\mathbf n}s}$ of the sources given the binned $uv$ data
$\{\overline{V}_{c}\}$. (We will describe how the centroid and
shapelet scales are chosen in practice in \S\ref{sim}). In principle,
the full $uv$ plane provides complete shape information for the
sources. However, due to the finite number and non-uniform spacings
of the antennae, the $uv$ (Fourier) space is poorly sampled, thus
hampering the decomposition. This prevents us from performing a simple
linear decomposition as is done with optical images in real space (see
Paper I). This problem can be largely resolved by making a linear fit
to the $uv$ plane with the shapelet coefficients as the free
parameters. 

For this purpose, the first step is to compute the binned visibilities
$\overline{V}_{c}^{{\mathbf n}s}$ corresponding to each shapelet basis
functions $B_{{\mathbf n}s}({\mathbf l}-{\mathbf l}_{s};\beta_{s})$
for each source $s$. This can be done by first computing the time- and
frequency-averaged visibility $\overline{V}_{ij}^{{\mathbf n}s}$ by
setting $f({\mathbf l})=B_{{\mathbf n}s}({\mathbf l}-{\mathbf
l}_{s};\beta_{s})$ in Equations~(\ref{eq:v_ij}) and
(\ref{eq:v_ij_bar}). To prevent potential biases introduced by the
binning scheme, we evaluate the basis functions at every visibility
point and then average them inside each cell to compute
$\overline{V}_{c}^{{\mathbf n}s}$ just as in
Equation~(\ref{eq:v_c}). Note that this ensures that the systematic
distortions induced by the primary beam, bandwidth smearing,
time-averaging and non-coplanarity can all be fully corrected in our
method.

The next step is to form and minimize
\begin{equation}
\label{eq:chi2}
{\chi}^{2}=({\mathbf d} - {\mathbf M ~f})^{T}~{\mathbf
C}^{-1}~({\mathbf d} - {\mathbf M~f}),
\end{equation}
where ${\mathbf d}=\{ \overline{V}_{c} \}$ is the data vector,
${\mathbf M}=\{ \overline{V}_{c}^{{\mathbf n}s} \}$ is the theory matrix,
and ${\mathbf f}=\{ f_{{\mathbf n}s} \}$ is the parameter vector. The
covariance error matrix 
\begin{equation}
{\mathbf C} = {\rm cov}[{\mathbf d},{\mathbf d}] =
\left\langle ({\mathbf d} - \langle
{\mathbf d} \rangle)^{T}{(\mathbf d} - \langle {\mathbf d} \rangle)
\right\rangle
\end{equation}
for the binned visibilities can be estimated in practice either from
the distribution of the visibilities in each bin or from the error
tables provided by the interferometric hardware.

Because the model is linear in the fitting parameters, the best-fit
parameters $\widehat{\mathbf f}$ can be computed analytically as
(e.g., Lupton 1993)
\begin{equation}
\label{eq:f}
\widehat{\mathbf f}= ({\mathbf M}^{T}{\mathbf C}^{-1}{\mathbf
M})^{-1}{\mathbf M}^{T}{\mathbf C}^{-1}{\mathbf d}.
\end{equation}
The covariance error matrix ${\mathbf W}={\rm cov}[\widehat{\mathbf
f},\widehat{\mathbf f}]$ of the best-fit parameters is given by
\begin{equation}
\label{eq:cova}
{\mathbf W}= ( {\mathbf M}^{T} {\mathbf C}^{-1} {\mathbf M} )^{-1}
\end{equation}
This provides us with an estimate for the shapelet coefficients for
each source and for their full covariance matrix. Note that all
sources are fitted simultaneously thus avoiding (or at least
quantifying) potential cross-talk between different sources (e.g.,
when a sidelobe from one source falls at the position of a second
source). The coefficient covariance matrix can also be used to
determine degeneracies produce by the finite $uv$ sampling of the array.

\subsection{Combining the Shapelet Coefficients}
\label{weighting}
Now that we have derived estimates $\widehat{f}_{\mathbf n}$ for the
shapelet coefficients $f_{\mathbf n}$ for each source in a pointing,
we can combine them to construct an image and to compute useful
quantities such as the fluxes, centroids and sizes of the sources.

We first consider the practical problem of co-adding several pointings
to derive an optimal image of a source. Let $\widehat{f}_{{\mathbf
n}p}$ be the coefficients of a source derived from pointing $p$, and
let $W_{{\mathbf nm}p}$ be the associated covariance error matrix
(from Eq. (\ref{eq:cova})). It is easy to see that the error in the
co-added coefficients $\widehat{f}_{\mathbf n}$ will be minimized if
they are given by the weighted sum
\begin{equation}
\widehat{f}_{\mathbf n} = \frac{ \sum_{p} W_{{\mathbf
n}\mathbf{n}p}^{-1}\widehat{f}_{{\mathbf n}p}}{\sum_{p} 
W_{{\mathbf n}{\mathbf n}p}^{-1}}.
\end{equation}
The covariance error matrix $W_{{\mathbf n}\mathbf{m}} =
{\rm cov}[ \widehat{f}_{\mathbf n},\widehat{f}_{\mathbf m}]$
of the co-added coefficients are then given by
\begin{equation}
W_{{\mathbf n}{\mathbf m}} = \frac{ \sum_{p} 
W_{{\mathbf n}{\mathbf n}p}^{-1} 
W_{{\mathbf n}{\mathbf m}p}  
W_{{\mathbf m}{\mathbf m}p}^{-1}} 
{ \left(\sum_{p} W_{{\mathbf n}{\mathbf n}p}^{-1} \right)
\left(\sum_{p} W_{{\mathbf m}{\mathbf m}p}^{-1} \right)}.
\end{equation}

\begin{figure*}[t]
\centering
\includegraphics[width=6.in]{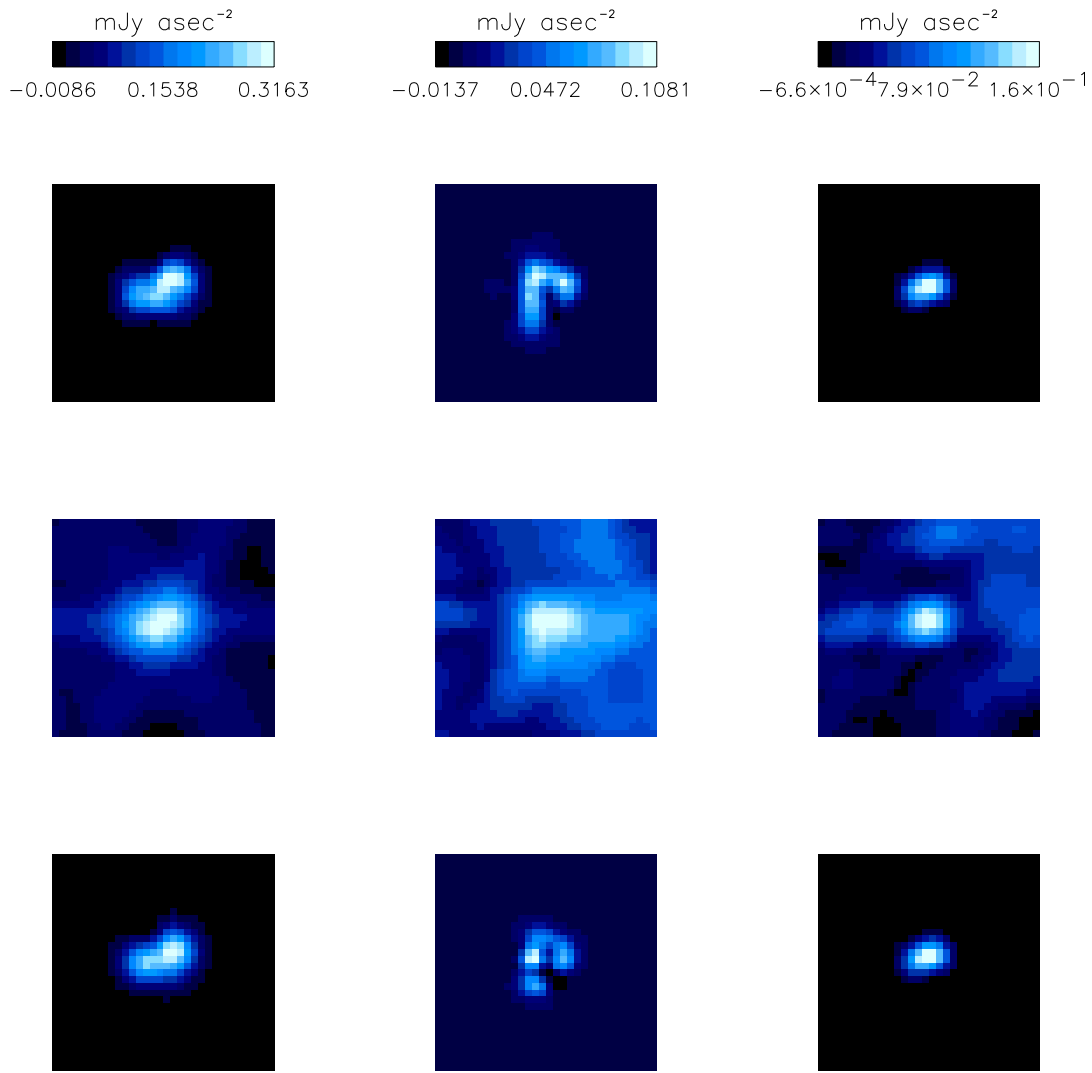}
\figcaption{Three sources from the simulation of a FIRST grid
pointing. The input images (before the addition of noise), dirty
images, and shapelet reconstructions are shown from top to bottom,
respectively.  The images are 32$''$ across with a 1$''$ pixel size,
and the resolution is $\sim$ 5$''$.4 (FWHM).  The input source flux
densities are (16.0, 4.3, 3.7) mJy from left to right, and the
recovered shapelet flux densities calculated using
Equation~(\ref{eq:flux}) are (16.1, 4.1, 3.8) mJy, respectively, where
(10, 21, 6) shapelet coefficients were used in the reconstruction. The
noise level in the simulation is about 0.3 mJy beam$^{-1}$.
\label{fig:sim}}
\end{figure*}

We can then find an optimal weighting to reconstruct the image of a
source from the estimated coefficients $\widehat{f}_{\mathbf n}$. To
do so we seek the reconstructed coefficients given by
\begin{equation}
f^{r}_{\mathbf n} = \phi_{\mathbf n} \widehat{f}_{\mathbf n}.
\end{equation}
The weights $\phi_{\mathbf n}$ are chosen so that
the reconstructed image $f^{r}({\mathbf l}) = \sum_{\mathbf n} f^{r}_{\mathbf
n} B_{\mathbf n}({\mathbf l})$ is `as close as possible' to the true image
$f({\mathbf l})$, in the sense that the least-square difference
\begin{equation}
\int d^{2}l~\left[ f^{r}({\mathbf l})-f({\mathbf l}) \right]^2 =
   \sum_{\mathbf n} [ f^{r}_{\mathbf n} - f_{\mathbf n} ]^{2}
\end{equation}
is minimized. It is easy to show that this will be the case when
\begin{equation}
\phi_{\mathbf n} = \frac{ |f_{\mathbf n}|^2 }{ |f_{\mathbf n}|^2 +
W_{{\mathbf n}{\mathbf n}}} \approx \frac{ |\widehat{f}_{\mathbf n}|^2 -
W_{{\mathbf n}{\mathbf n}}}{|\widehat{f}_{\mathbf n}|^2},
\end{equation}
where the right-hand side provides an approximation which can be
directly derived from the data. This weighting amounts to Wiener
filtering in Shapelet space, in analogy with that performed in Fourier
Space (see, e.g., Press et al. 1987). Figures \ref{fig:sim} and
\ref{fig:data} show several reconstructed images using this weighting
scheme. Note that this produces an estimate for the {\em deconvolved}
image of the source. For display purposes, it is sometimes useful to
smooth the reconstructed image by a Gaussian kernel (the 'restoring
beam' in radio parlance). This can easily be done in shapelet space by
multiplying the coefficients by the analytic smoothing matrix
described in Paper I.

While Wiener filtering yields an optimal image reconstruction, it is
{\it not} to be used to measure source parameters such as flux, centroid,
size, etc. Instead, an unbiased estimator for shape parameters can
be derived directly from the shapelet coefficients (see Paper I).
For instance, an estimate for the flux $F\equiv \int d^{2}l f({\mathbf
l})$ of a source is given by $\widehat{F} =
{\mathbf A}^{T} \widehat{\mathbf f}$ where
\begin{equation}
\label{eq:flux}
A_{n_{1}n_{2}} = \pi^{\frac{1}{2}} \beta 
2^{\frac{1}{2}(2-n_{1}-n_{2})} 
\left( \begin{array}{c} n_{1} \\ n_{1}/2 \end{array} \right)^{\frac{1}{2}}
\left( \begin{array}{c} n_{2} \\ n_{2}/2 \end{array}
\right)^{\frac{1}{2}},
\end{equation}
if $n_{1}$ and $n_{2}$ are both even (and vanishes otherwise). The
variance uncertainty in the flux is then simply
\begin{equation}
\label{eq:error}
\sigma^{2}[\widehat{F}] = {\mathbf A}^{T} {\mathbf W} {\mathbf A},
\end{equation}
which provides a robust estimate of the signal-to-noise
SNR$=\widehat{F}/\sigma[\widehat{F}]$ of the source.  Similar
expressions can be used to compute the centroid and rms size of the
source. This can be easily generalized to compute in addition the
major and minor axes of the source and its position angle. Note that
these expressions are, again, estimates for deconvolved quantities.

\section{Test of the Method}
\label{results}
As an application, we consider the FIRST radio survey (Becker et
al. 1995; White et al. 1997), being conducted with the VLA at 1.4
GHz in the B configuration. For this survey, the primary beam FWHM is
$\sim$ 30$'$ and the angular resolution is $5''.4$ (FWHM).  The survey
currently contains about $7.2\times 10^5$ sources with a $5\sigma$
flux-density limit of 1.0 mJy over $A \simeq 8,000$ deg$^2$; the mean source
redshift is $\langle z \rangle \sim 1$. Observing time has been
allocated to extend its coverage to 9,000 deg$^2$.  The survey is
composed of 165-second `grid-pointings' with a time-averaging interval
$\Delta t = 5$ seconds.  It was conducted in the spectral synthesis
mode, with a channel bandwidth of $\Delta \nu =$ 3 MHz. Because this
wide-field survey was performed in the snapshot mode, its $uv$
sampling is very sparse. This makes shape reconstruction particularly
challenging for FIRST, providing a good test for our method.

As explained in \S\ref{interferometers}, higher order effects such as
bandwidth smearing and time-averaging produce small distortions in the
reconstructed image shapes if they are left uncounted for. These must
be carefully corrected for high-precision statistical measurements of
object shapes such as those required in weak lensing surveys. The
effects are, however, very small and not noticeable on an
object-by-object basis.  For the purpose of this test, we thus ignore
these effects and instead focus on the dominant factor in shape
reconstruction, the finite and discrete $uv$ sampling.

\subsection{Simulations}
\label{sim}
As a first test, we generated simulated VLA data using the
observational parameters of FIRST. A grid pointing was generated at
zenith with 33 5-second time-averaging intervals and 14 3-MHz channels
in the B configuration.  Simulated sources were randomly distributed
within 23$'$.5 of the phase center, the cutoff adopted for creating
the final co-added FIRST maps (Becker et al. 1995); the number
density, flux density and size distributions chosen for the sources
were similar to sources in the FIRST catalog. After generating the
visibilities, we added uncorrelated Gaussian noise to the real and
imaginary component of each $uv$ data point, with an rms of $\sigma_v
= \sigma_n N_{\rm vis}^{0.5}$, where $N_{\rm vis}$ is the total number
of visibilities. The real-space rms noise $\sigma_n$ was set to $0.3$
mJy beam$^{-1}$, which is somewhat higher than the typical FIRST noise
level, $\sim$ 0.2 mJy beam$^{-1}$.
  
We then simultaneously fitted all 23 sources in the grid pointing
directly in the $uv$ plane. We imposed the constraint that that source
intensities are real (i.e., non-imaginary). Each source $s$ was
modeled as a shapelet with scale $\beta_{s}$, maximum shapelet order
$n_{{\rm max},s}$, and center position ${\mathbf l}_{s}$.  In
principle, it is possible to determine these parameters with a source
detection algorithm which directly uses shapelets. One can, for
instance, tile ground-state shapelets with different sizes in the $uv$
plane, and thus detect sources with different sizes.  However, this is
computationally expensive and, since the FIRST catalog is conveniently
available, we have not implemented this algorithm.

Instead, good choices for these parameters were derived from the FIRST
catalog, which lists basic shape parameters for each source, such as
its centroid, flux density, major and minor axes, and position angle,
all obtained from an elliptical Gaussian fit. The shapelet position
${\mathbf l}_{s}$ was simply set to the centroid position from the
catalog. The choices for the shapelet scales $\beta_{s}$ and maximum
shapelet orders $n_{{\rm max},s}$ were derived as follows. As
described in Paper I, the Hermite basis functions have two natural
scales: $\theta_{\rm max}$ corresponding to the overall extent of the
basis functions, and $\theta_{\rm min}$ corresponding to the
smallest-scale oscillations in the basis functions. These scales are
related to the shapelet scale and maximum order by $\theta_{\rm max}
\sim \beta (n_{\rm max}+1)$ and $\theta_{\rm min} \sim \beta (n_{\rm
max}+1)^{-1}$. As $n_{\rm max}$ increases, the large-scale size of
the shapelet grows, while its small-scale features become finer. The
shapelet thus becomes more extended both in real and in Fourier
space. We therefore choose $\theta_{\rm max}$ to be the rms major axis
from the FIRST catalog, and $\theta_{\rm min}$ to correspond to the
longest baseline of the VLA: $\sim$ 1$''$.8 (rms) in real space. This
provides us with a choice for $\beta \simeq (\theta_{\rm
max}~\theta_{\rm min})^{0.5}$ and for $n_{\rm max} \simeq
\frac{\theta_{\rm max}}{\theta_{\rm min}}-1$ for each source.
 
Solving Equation~(\ref{eq:chi2}), we then obtain the shapelet
coefficients and the covariance matrix using Equations~(\ref{eq:f})
and (\ref{eq:cova}).  The results are presented in Fig.~\ref{fig:sim},
where the input images (before the addition of noise), inverse
Fourier-transformed $uv$ data (`dirty' images), and
shapelet-reconstructed images (with Weiner filtering, see
\S\ref{weighting}) of three of the sources are shown.  Each image
shown is 32$''$ across and the resolution is about 5$''$.4 (FWHM). The
poor $uv$ sampling of FIRST and the effect of noise are evident in the
dirty images.  For both resolved (left panels) and unresolved or
marginally resolved (right panels) sources, the reconstructions agree
with the inputs very well.  The more complicated structure in the
central panel is not fully recovered by shapelets. This is
expected, since the small-scale structure of the source is not resolved
and therefore can not be fully recovered in the reconstruction.

The comparison between the input and shapelet-reconstructed flux
density for all sources in the grid pointing is shown in
Figure~\ref{fig:flux}. The shapelet flux density is given by
Equation~(\ref{eq:flux}) and its 1$\sigma$ error by
Equation~(\ref{eq:error}). The source flux densities are
well-recovered by the shapelets in an unbiased manner. Note the range
of error bars at a given input flux is due to the range of source
sizes. For instance, for an input flux density of about 2 mJy, the
source with a relatively large error bar has a major axis of about
$8''$ (FWHM), while those with small error bars are unresolved or
barely resolved (major axis FWHM $< 5''$). In general, we find the
shapelet reconstruction from the sparsely sampled and noisy simulated
data to be in good agreement with the input (noise-free) image.  Note
that our method can be used to identify and discard spurious sources
arising from sidelobes and other artifacts in the dirty image. Indeed,
when we place an extra shapelet centered at a random positions in the
field, the coefficients of that shapelet are consistent with zero.

\vskip0.1in
\includegraphics[width=3.in]{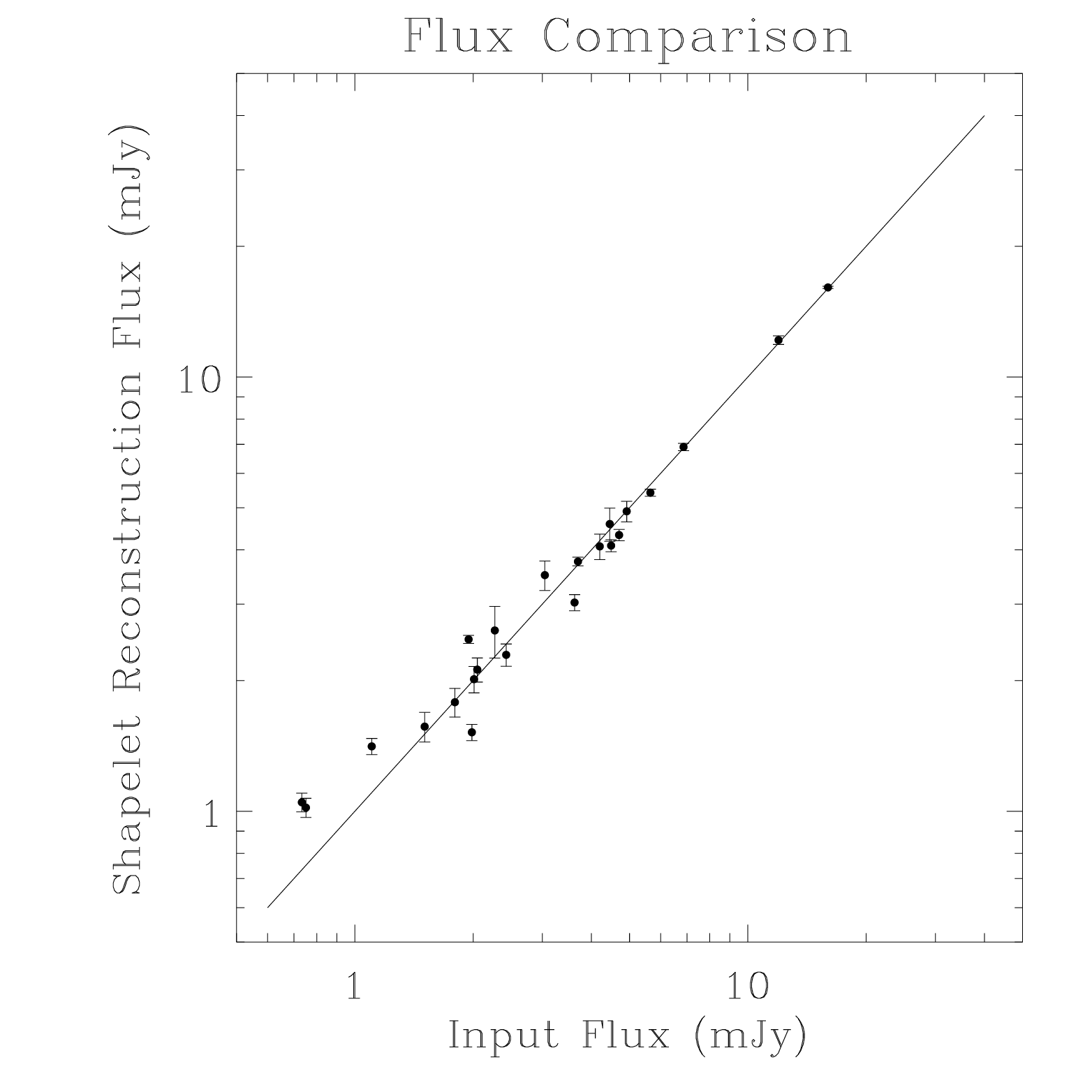}
\figcaption{Flux density comparison for the
simulations: the input flux densities of the 23 sources in the
simulated pointing are compared
with the shapelet reconstructed flux densities.  The solid line
corresponds to a perfect reconstruction-- i.e., to a recovered flux
density equal to the input flux density.
\label{fig:flux}}
\vskip0.1in

\subsection{Data}
Next, we test our method by applying it to one of the FIRST grid
pointings (14195+38531). For this purpose, we selected all sources
within 23$'$.5 of the phase center from the FIRST catalog with a
measured flux density limit (i.e., including the primary beam
response) of 0.75 mJy\footnote{As explained in Becker et al. (1995), a
map flux density of 0.75 mJy corresponds to a source flux density of
1.0 mJy owning to ``CLEAN bias'' corrections.}.  For each of the
resulting 23 sources, we use the source major axis to estimate $\beta$
and $n_{\rm max}$ as described in the previous section.  We then
simultaneously fit all the sources for the shapelet coefficients
directly in the $uv$ plane.  Note that the shapelet coefficients
obtained are deconvolved coefficients.

\begin{figure*}[t]
\centering
\includegraphics[width=6in]{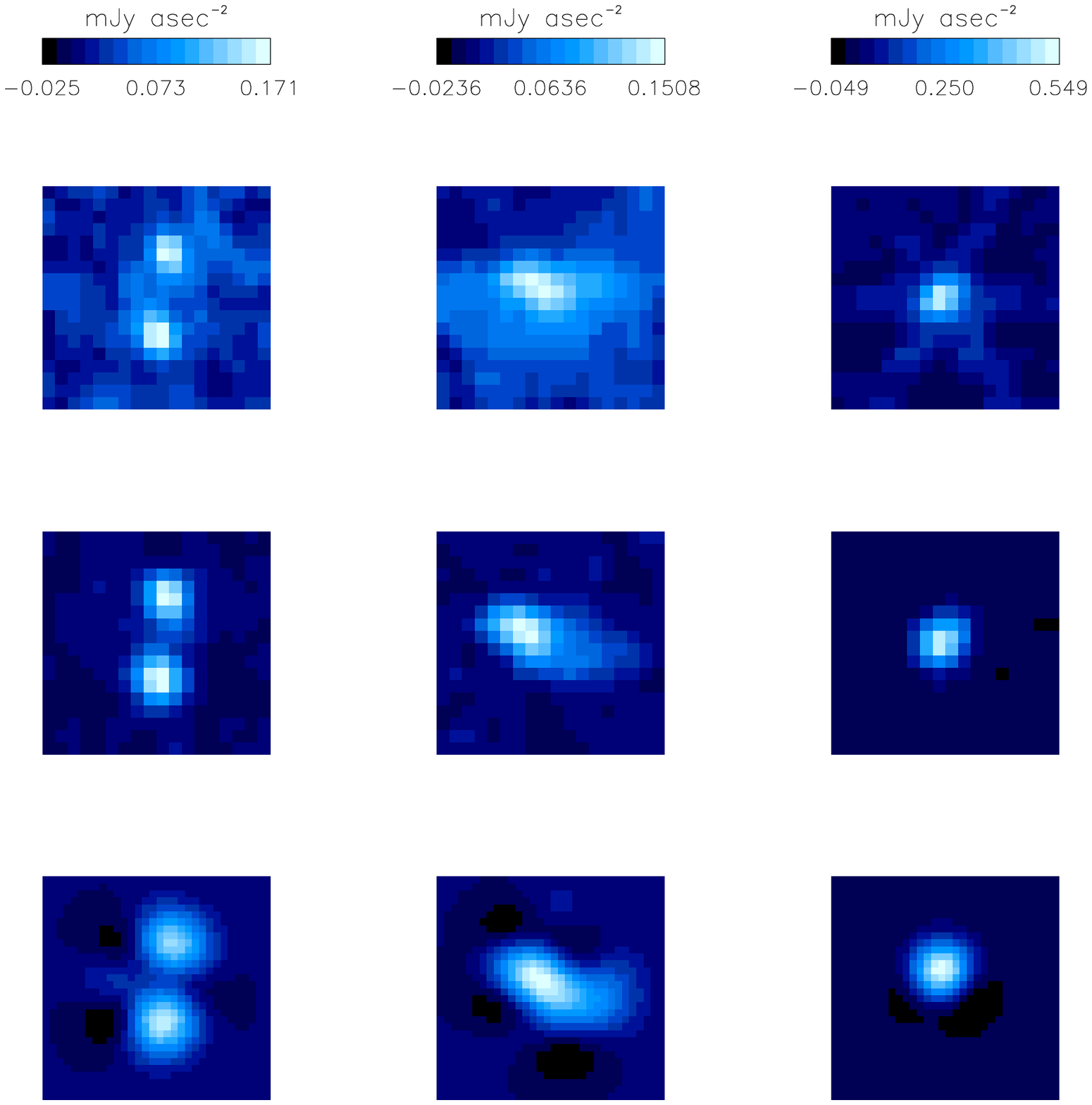}
\figcaption{Three sources from one of the FIRST grid
pointings.  The dirty images, CLEAN images, and shapelet
reconstructions are shown from top to bottom, respectively.  The
images are 32$''$ across and the resolution is about 5$''$.4 (FWHM).
The dirty and CLEAN images are displayed with a 1$''$.8 pixel size,
while the shapelet reconstruction images have 1$''$ pixels. The source
flux densities measured by integrating fitted elliptical Gaussians to
CLEANed sources are (13.4, 14.2, 19.5) mJy, from left to right, and
the recovered shapelet flux densities calculated using
Eq. (\ref{eq:flux}) are (17.0, 13.8, 22.5) mJy, respectively. We used
(15, 28, 6) shapelet coefficients in the reconstructions, along with
Wiener filtering and smoothing by a Gaussian restoring beam with a
standard deviation of 2.3$''$.
\label{fig:data}}
\end{figure*}

Figure~\ref{fig:data} shows the reconstruction of three representative
sources in the bottom panels. Also shown for comparison are the images
of the sources constructed using the standard AIPS CLEAN algorithm
with a CLEANing limit of 0.5 mJy (central panel), along with the dirty
images (top panel).  Each panel is 32$''$ across and the FWHM of the
FIRST resolution is 5$''$.4.  The shapelet method does not involve
image pixels in the modeling; one is therefore free to specify the
pixel size when constructing the images.  Here the dirty and CLEANed
images have pixel sizes of 1$''$.8, while the shapelet images have
pixel sizes of 1$''$ and thus show finer details.  For demonstration,
the shapelet reconstructions have been Weiner-filtered using the
resulting covariance matrix. For a direct comparison, they have also
been smoothed with a Gaussian kernel with a standard deviation of
2$''$.3, reproducing the 'restoring beam' of the CLEANed image. We
find that the shapelet reconstructions compare well with the CLEANed
images. 

In further tests, we have encountered cases where a bright source ($>
100$ mJy) lies in or near a grid pointing. We have found that the
presence of the bright source does not affect the fit of the other
sources in the grid in a noticeable way.  Our method can thus well
handle the dynamical range of the FIRST survey, which spans more than
3-orders of magnitude. For fainter sources ($< 1$ mJy; i.e., detection
SNR $< 6$), the reconstructions are rather poor at times, in contrast
to those for brighter sources (which are almost always well
fitted). This is of course reasonable, given the larger impact of
noise for faint sources.

In Figure~\ref{fig:cova} we display a portion of the covariance matrix
for the shapelet coefficients for the nine sources in the pointing
with the highest peak flux densities.  The horizontal and vertical
lines separate the nine sources.  The diagonal line from the
lower-left to the upper-right corner represents the variance of the
shapelet coefficients. The block-diagonal boxes are the covariance
matrix of the coefficients of the nine sources. The off-diagonal
blocks quantify the cross-talk between sources. Note that the
correlation between coefficients are roughly an order of magnitude
smaller than the variance. Figure~\ref{fig:cova_s4} shows the error in
the shapelet coefficients (n1,n2) of the source shown in the left
panels of Figure~\ref{fig:data}. (These errors are the diagonal
segment of the 4$^{th}$ diagonal box counting from the lower left in
Fig.~\ref{fig:cova}). In general, we find that higher-$n$ coefficients
tend to be noisier. This is expected since convolution (or, 
equivalently, $uv$ sampling) suppresses the small scale information
encoded by coefficients with large $n$ (see paper I). The covariance
matrix thus provides us with useful information on the error in
each coefficient, and quantifies cross-talk between coefficients both
within and among sources.

\subsection{Computation}
Since the shapelet coefficients of all sources are simultaneously fit
to a large number of visibilities, the computing memory required for
the calculation is not negligible. We have implemented the method on
the UK COSMOS SGI Origin 2000 supercomputer, which has 64 R10000 MIPS
processors with a shared-memory structure.  Numerically, the shapelet
coefficients can be obtained by performing simple matrix operations as
in Equation~(\ref{eq:f}), or by solving the linear least-squares
problem, ${\mathbf M ~f} = {\mathbf d}$, using matrix factorization or
singular value decomposition, and assuming that the data covariance
matrix ${\mathbf C}$ is diagonal. Both methods can be efficiently
parallelized. With our binning scheme, the run-time memory required
for this particular FIRST grid pointing was about 700 MB, for 23
sources and a total of 177 shapelet parameters. The CPU time required
was about 26 seconds with 10 processors or about 5 minutes in scalar
mode.  For other grid pointings with different numbers of sources, the
computation time ranges between 20 and 60 seconds with 10 processors,
with a run time memory between 0.5 to 1.5 GB.

\section{Applications to Weak Lensing}
\label{lensing}
Weak gravitational lensing is now established as a powerful method for
mapping the distribution of the total mass in the Universe (for
reviews see Mellier 1999; Bartelmann \& Schneider 2000). This
technique is now routinely used to study the dark matter distribution
of galaxy clusters and has recently been detected in the field
(Wittman et al 2000; van Waerbeke et al 2000; Bacon, Refregier \&
Ellis 2000; Kaiser et al 2000; Maoli et al 2001; Rhodes, Refregier \&
Groth 2001; van Waerbeke et al 2001).  All studies of weak lensing
have been performed in the optical and IR bands, where the images are
directly obtained in real space.  

\includegraphics[width=3.in]{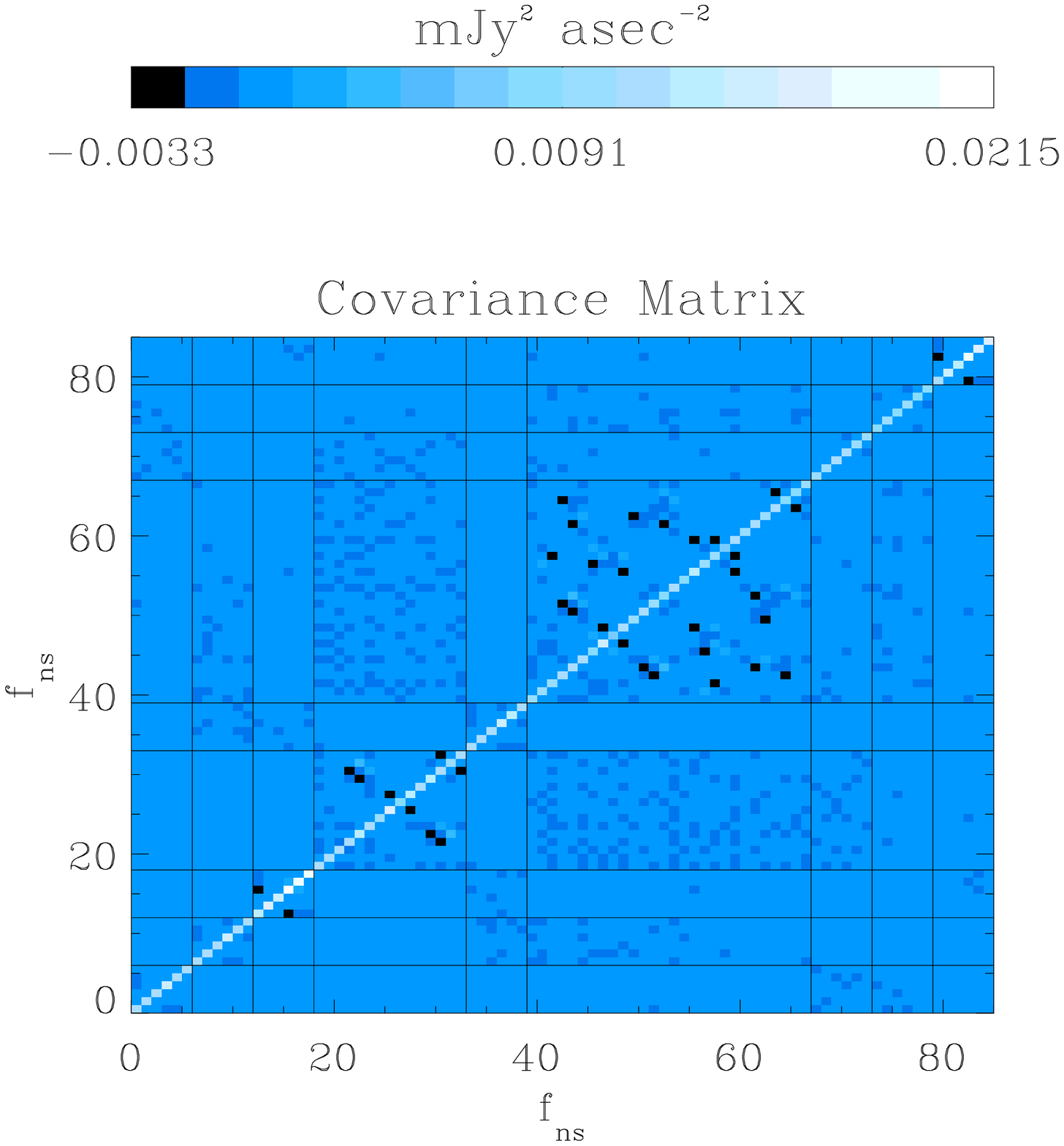}
\figcaption{The covariance matrix for the shapelet
coefficients of nine sources in the FIRST grid pointing.  The sources
are separated by the horizontal and vertical lines. The diagonal
entries correspond to the variance (i.e., error) of each
shapelet coefficient. The block-diagonal boxes are the covariance
matrix of the coefficients for each of the nine sources. The
off-diagonal boxes quantify the cross-talk between sources.
\label{fig:cova}}
\vskip0.1in

\includegraphics[width=3.in]{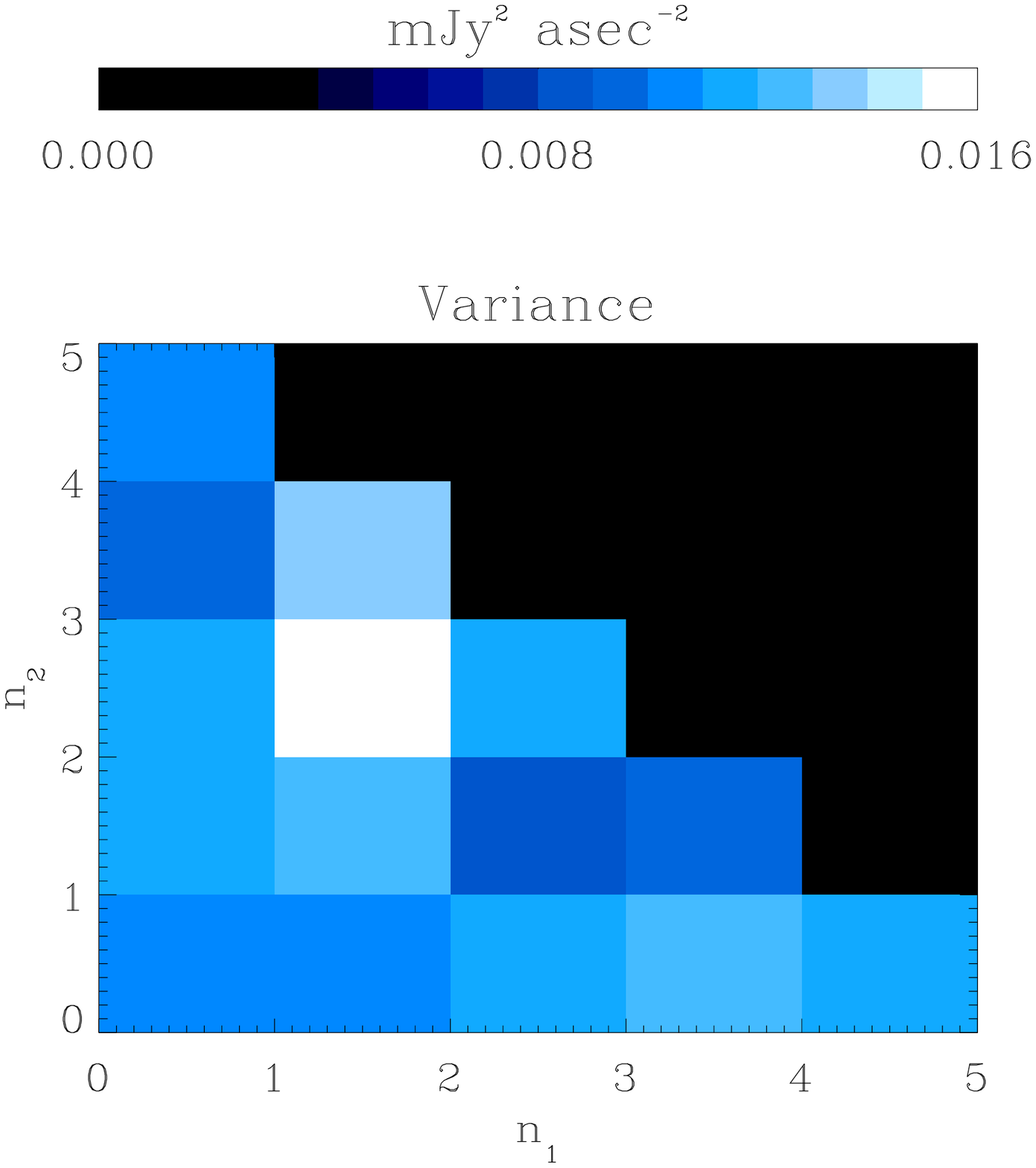}
\figcaption{The error matrix for one of the FIRST sources plotted in the (n1,
n2) plane. The shapelet reconstructed image of this source is shown on the
bottom-left panel of Fig.\ref{fig:data}. These variances can also be
used to Weiner filter the coefficients for image reconstruction, as
described in \S\ref{weighting}.
\label{fig:cova_s4}}
\vskip0.2in

There are a number of reasons to try to extend these studies to
interferometric images in the radio band. Firstly, the brightest radio
sources are at high redshift, thereby increasing the strength of the
lensing signal. Secondly, radio interferometers have a well-known and
deterministic convolution beam, and thus do not suffer from the
irreproducible effects of atmospheric seeing.  Thirdly, existing
surveys such as the FIRST radio survey (Becker et al. 1995; White el
al. 1997) provide a sparsely sampled but very wide-area survey, which
offers the unique opportunity to measure a weak lensing signal on
large angular scales (Kamionkowski et al. 1998; Refregier et al. 1998;
see also Schneider (1999) for the case of SKA).  Finally, surveys at
higher frequencies or in more extended antenna configurations could
potentially yield very high angular resolution and are not limited by
the irreducible effects of the seeing disk in ground-based
optical surveys.

Because the distortions induced by lensing are only on the order of
1\%, the shapes of background objects must be measured with high
precision. In addition, systematic effects such as the convolution
beam and instrumental distortions must be tightly controlled. For this
purpose, a number of shear measurement methods have been
developed. The original method of Kaiser, Squires \& Broadhurst (1995)
was found to be acceptable for current cluster and large-scale
structure surveys (Bacon et al. 2000b; Erben et al. 2000), but are not
sufficiently reliable for future high-precision surveys. Consequently,
several other methods have been proposed (Kuijken 1999; Kaiser 2000;
Rhodes, Refregier \& Groth 2000, Berstein \& Jarvis 2001).

Recently, Refregier \& Bacon (2001, Paper II) developed a new method
based on shapelets and demonstrated its simplicity and accuracy for
ground-based surveys. It is thus straightforward to apply this method
to interferometric measurements. Indeed, the shapelet coefficients
$f_{\mathbf n}$ which we derive from the fit on the $uv$ plane (after
co-adding if required) are already deconvolved from the effective beam
and can thus be directly used to estimate the shear. This can be done
using the estimators for the shear components $\gamma_{1}$ and
$\gamma_{2}$ which are given by (see Paper II)
\begin{equation}
\gamma_{i} = \frac{ f_{\mathbf n} - \langle f_{\mathbf n}
\rangle}{\sum_{\mathbf m} 
\hat{S}_{i{\mathbf n}{\mathbf m}} \langle f_{\mathbf m} \rangle },
\end{equation}
where the sum is over even (odd) shapelet coefficients for
$\gamma_{1}$ ($\gamma_{2}$) and the brackets denote an average over an
(unlensed) object ensemble. The matrix $\hat{S}_{i{\mathbf n}{\mathbf
m}}$ is the shear matrix, and can be expressed as simple combinations
of ladder operators in the QHO formalism. These estimators for
individual shapelet components are then optimally weighted and
combined to provide a minimum-variance estimator for the shear. This
permits us to achieve the highest possible sensitivity (while
remaining linear in the surface brightness) by using all the available
shape information of the lensed sources.

In Kamionkowski et al. (1998) and Refregier et al. (1998), it has been
shown that the FIRST radio survey is a unique database for measuring
weak lensing by large-scale structure on large angular scales. In a
future paper, we will apply the method described here to this survey,
search for the lensing signal, and, from its amplitude, derive
constraints on cosmological parameters.

\section{Conclusions} 
\label{conclusion} 
We have presented a new method for image reconstruction from
interferometers. Our method is based on shapelet decomposition and is
simple and robust. It consists of a linear fit of the shapelet
coefficients directly in the $uv$ plane, and thus permits a full
correction of systematic shape distortions caused by bandwidth
smearing, time-averaging and non-coplanarity.  Because the fit is
linear in the shapelet coefficients it can be implemented as simple
matrix multiplications. It provides the full covariance matrix of the
shapelet coefficients which can then be used to estimate errors and
cross-talk in the recovered shapes of sources. We have shown how
source shapes from different pointings can be easily co-added using a
weighted sum of the recovered shapelet coefficients. We have also described
how the shapelet parameters could be combined to derive optimal image
reconstruction, photometry, astrometry and pointing co-addition.

Our method can be efficiently implemented on parallel computers. We
find that a fit to all the sources in a FIRST grid pointing takes
about 1 minute on an Origin 2000 supercomputer with 10 processors (10
minutes in scalar mode). Because we are fitting all sources
simultaneously, 0.5 to 1.5 GB of memory is required.

To test our methods, we considered the observing conditions of the
FIRST radio survey (Becker et al. 1995; White et al. 1997) whose
snapshot mode yields a sparse sampling in $uv$ space.  Using
numerical simulations tuned to reproduce the conditions of FIRST, we
find that the sources are well-reconstructed with our method. We have
also applied our method to a FIRST snapshot pointing and found that
the shapes are well-recovered. The reconstruction of our method
compares well with the CLEAN reconstruction, without suffering the
potential biases inherent in the latter method. Our method is thus
well-suited for applications requiring quantitative and high-precision
shape measurements.

In particular, our method is ideal for the measurement of the small 
distortions induced by gravitational lensing in the shape of
background sources by intervening structures. Such a measurement from
CLEANed images may well not be practical since the systematic
distortions induced by that method are very difficult to
control. (One could perhaps imagine running numerical simulations to
calibrate the shear estimator, but this would be both computationally
expensive and rather uncertain). We have shown how our results can be
combined with the shear measurement method described in Refregier \&
Bacon (2001) to derive a measurement of weak lensing with
interferometers. This is facilitated both by the fact that our
recovered shapelet coefficients are already deconvolved from the
effective (dirty) beam, and as a consequence of the remarkable
properties of shapelets under shears.

Our method therefore opens the possibility of high-precision
measurements of weak lensing with interferometers. While to date all
weak-lensing studies have been carried using optical data (and
therefore in real space), an interferometric measurement of weak
lensing in the radio band is very attractive (Kamionkowski et
al. 1998; Refregier et al. 1998; Schneider 1999). Indeed, the lensing
signal is expected to be larger because radio sources have a higher
mean redshift. In addition, such a measurement would not suffer from
the irreproducible effects of atmospheric seeing. Instead, the
effective (dirty) beam is fully known for interferometers and the
noise properties of the antennas are well-understood. As a result, the
impact of systematic effects, the crucial limitation in the search for
weak lensing, are expected to be lower with radio interferometers. In
a future paper, we will describe our measurement of weak lensing by
large-scale structure with the FIRST survey using the present method.

\acknowledgements{We thank David Bacon, David Helfand, Jacqueline van
Gorkom, Ue-Li Pen, Rick Perley and Marc Verheijen for useful
discussions. At Columbia, this work was supported by NSF grant
AST-98-0273. AR was supported by the EEC TMR network on Gravitational
Lensing and by a Wolfson College Research Fellowship. This work was
performed on the UK-CCC COSMOS facility, which is supported by HEFCE
and PPARC and conducted in cooperation with Silicon Graphics/Cray
Research utilizing the Origin 2000 supercomputer.}

\end{document}